\documentclass[cits]{PoS}

\title{Open issues in confinement, for the lattice and for center vortices}

\ShortTitle{Open issues in confinement }

\author{\speaker{John Cornwall}%
        University of California at Los Angeles (UCLA)\\
        E-mail: \email{cornwall@physics.ucla.edu}}

\abstract{Center vortices have been around for more than thirty years,
well-confirmed on the lattice, and very successful in explaining the basics
of confinement, yet there are still open questions unstudied either on the
lattice or in theory.  The first is that basic confinement in the center
vortex picture is {\em topological}, coming from gluonic solitons where the
gluons have no direct coupling to the Wilson loop, and makes no reference
to any
particular surface (whose area would appear in the area law) or fluctuation
dynamics of this surface.  Only in $d=2$ (flat Wilson loops) is it obvious
what surface must be involved, and in this dimension there is no room for
fluctuations.  This makes it hard to understand the L\"uscher term and
other properties of the fluctuating confinement surface for $d>2$.  I make
the obvious, but unconfirmed to date, conjecture that   in topological
confinement for {\em non-planar} Wilson loops the area law is the
exponential of a string tension times the area of a {\em minimal} surface
spanning the Wilson loop.  Less obvious is whether, in this purely
topological picture, this minimal surface shows the correct L\"uscher term,
or whether this term must come from gluons propagating between points on
the Wilson loop (as possibly described by fishnet graphs and their relative
the gluon chain model).  
Closely-related issues are the structure of the area law for {\em two}
coaxial Wilson loops, as the distance between them along the axis grows;
the resulting Casimir force between hadrons; and the behavior of $k$-string
tensions for $SU(N)$ with $N>3$. I suggest a program of both lattice and
theoretical studies, focused on center vortices and the pinch technique, to
explore these and other issues:  1)  Calculate the area law and its
fluctuations for {\em
non-planar} Wilson loops, or for pairs of flat Wilson loops, in a
center-vortex-like ground state with a gas of vortices, but with no
gluon-Wilson loop coupling.      2)  Study more closely a picture I  outline
here of
reconciling center vortices and minimal surfaces with fishnet graphs and
the gluon-chain model, with the key ingredient of dynamically-massive
gluons.   3) Extend beyond perturbation theory the
old lattice work of Dashen and Gross on   background-field Feynman gauge
fixing  to extract the {\em gauge-invariant} off-shell Green's functions of
the pinch technique.       }

\FullConference{International Workshop on QCD Greens Functions, Confinement
and Phenomenology\\
                 September 7-11 , 2009\\
                 ECT Trento, Italy}

\begin{document}

 \section{Introduction}

Center vortices have been around so long \cite{thooft} that one might think
there is nothing new to say about them, and in fact most of the important
developments, both in theory and on the lattice, are covered in Greensite's
2003 review article \cite{green}.  The general view among many of us at
that time was that center vortices explained everything that needed
explaining,  although in certain cases the explanations might have been a
bit skimpy.  For example, center vortices explain the $N$-ality dependence
of confinement, including string breaking for $N$-ality zero
representations; the lack of confinement in the exceptional Lie groups
$G_2,F_4$, and $E_8$ that have trivial centers; along with nexuses
(monopole-like objects) they generate non-integral topological charge; and
are essential for chiral symmetry breakdown (CSB). But since 2003 many
workers have turned to other issues; in particular, on the lattice it has
become popular to study various gauge fixings, notably Landau and Coulomb
gauges.  

For center-vortex believers there are
important questions still awaiting convincing answers, including the details
of CSB and the $\theta$ dependence of quarkless $SU(N)$ gauge theories, 
but I will not discuss them here.  Instead, I want to raise what seems to
me to be an important point of principle so far not well-understood in QCD: 
What is a good first-principles
explanation of the L\"uscher term, given a confining condensate of center
vortices?  That is, how is a surface under tension formed spanning a Wilson
loop?  The reason this is not so well-understood is that it is easy to
understand confinement as a  purely {\em topological} phenomenon of the
linking of center vortices with a Wilson loop, but the basic topology makes
no reference to any  spanning surface or its area.  There are, of course,
candidate non-topological mechanisms for forming a surface under tension
that I (and Greensite, at this Workshop) will   discuss, but it is also
interesting to know whether or not a purely topological form of confinement
through an ensemble of center vortices actually leads to a minimal surface
with tension.  This is a challenging problem only for non-planar Wilson
loops, but it can be simulated much more easily than full QCD can be
simulated.

Aside from purely topological generation of a L\"uscher term, the other
candidates are gluon-chain models [Greensite, this Workshop;
\cite{tikt,greenthorn,greenole}] and fishnet graphs
\cite{nielole,sakvira,thooft2}.  I argue that there is a kind of $d=4$
duality between the chromomagnetic, source-free gluons bound up in the
closed two-surfaces of center-vortex solitons and gluons with
chromoelectric sources that propagate from one point on a Wilson loop to
another, while interacting with each other and forming a fishnet-like
two-surface (or, at fixed time, a gluon chain).  The chromoelectric gluons
in the gluon chain are analogous to, or dual to, the localized
monopole-like solitons \cite{ambgreen,corn122}, which I call nexuses, that
live on the center-vortex surface.   Entropic effects coming from the
coupling strength  ($4\pi /g^2\simeq 1$) distinguish between electric and
magnetic confinement.  This kind of approximate duality only makes sense if
the gluons have a dynamical mass $m$, which allows nexuses of mass $\sim
4\pi m/g^2$ to exist.

The final point of this talk concerns the greatly-increased activity in
lattice studies of off-shell Green's functions, where it has so far been
necessary to fix to a particular gauge (usually Landau or Coulomb).  This
is hardly in the spirit of the original Wilson formulation of gauge theory,
where one of the most important and basic features was gauge invariance. 
At this Workshop you will hear about recent advances in using the pinch
technique \cite{corna7,corna9,corn076}, which is a way of constructing {\em
gauge-invariant} off-shell Green's functions in the continuum.  My
colleagues Papavassiliou and Binosi, motivated by some one-loop
calculations of others, proved \cite{binpap} that to all orders the pinch
technique is exactly the same as the background-field Feynman gauge.  So my
plea to the lattice workers here:  Can you find a way of doing lattice
simulations in this gauge?  (Dashen and Gross \cite{dashg} did it long ago
to lowest order in the coupling, and Cucchieri {\em et al.} \cite{cucch}
have taken the first step toward an effective formulation of general
$R(\xi )$ gauges, where $\xi =1$ is the Feynman gauge.)

\section{Center vortices}

The first question to ask is whether there is a convincing model of center
vortices apart from the multiple lattice simulations (reviewed in
\cite{green}) showing that they exist.  Just after center vortices were
introduced by 't Hooft, I \cite{corn79} pointed out that a
dynamically-generated gluon mass made it easy to find center vortices as
solitons\footnote{In their simplest form, essentially the Abelian
Nielsen-Olesen vortex of the Abelian Higgs model, but in fact much more
complicated as we will
discuss.} of an infrared-effective action with a gauge-invariant mass term. 
  The Schwinger-Dyson
equations of the pinch technique \cite{corn076,binpap2,abp}  showed that
there was indeed a dynamical gluon mass of perhaps 600 MeV driven by
infrared slavery.  This result has been repeatedly confirmed on the lattice
(see the references in \cite{binpap2,abp}).   A dynamical gluon mass and  
the necessary long-range pure-gauge potentials are crucial ingredients in
the picture presented here.

I want to connect the L\"uscher term to (planar) fishnet graphs
\cite{nielole,sakvira,thooft2}.  Much of this early interest in fishnet
graphs centered on the possibility of finding in them the Veneziano
dual-resonance model.  This is not my concern here, which is to find a
description of a physical surface under tension.  It will be critical that
the gluons have a mass.  Before going into the fishnet/gluon chain models,
I will give a brief review of why the basic picture of center-vortex
confinement is {\em topological}.

\section{ Basic topological center-vortex confinement}

As far as confinement goes in QCD, there are two kinds of gluons.  The first
kind, type I (chromomagnetic), are the ones in the condensate of center
vortices; these source-free
gluons are parts of solitons that are only indirectly coupled to a Wilson
loop.  The second, type II (with chromoelectric sources such as quarks),
are the
ones that propagate from one point on a Wilson loop to another point,
interacting with other gluons as they go.  Correspondingly, 
there are two possibilities for forming a surface under tension, and I
suspect that both of them play an important role.  The first possibility is
that, for
deep mathematical reasons, the area in the area law for a condensate of
random mutually-avoiding vortices formed from type I gluons is the area of a
minimal surface and that this surface is under tension, for planar and
non-planar Wilson loops.  I suspect that this is so, and hope that the
relatively straightforward lattice simulations (because type II gluons can
be omitted) to investigate this question will soon be done. The second may
lie in some approximation to the properties of fishnet graphs.

Let me briefly review the essential ingredients of topological confinement
using a gauge-invariant massive Abelian model, suppressing all
irrelevancies.  For simplicity of
exposition I speak only of $d=3$ center vortices for a $U(1)$ subgroup of
gauge group $SU(2)$,
which are effectively Abelian closed $d=1$ loops, but everything goes
through in $d=4$ where vortices are closed $d=2$ surfaces.  This model has
long-range Goldstone-like fields that are necessary for gauge-invariant
gluon masses.  The action   is of London form:
\begin{equation}
\label{london}
I=\int d^3x \{\frac{1}{4}F_{ij}^2+\frac{m^2}{2}(A_i-\partial_i\phi)^2
 -i\oint_{\Gamma} d\tau \dot{z}_i\delta (x-z(\tau ))A_i\};
\end{equation}
the last term is the coupling to the current $J_i$ of the Wilson loop
$\Gamma$.

Type I center vortices are Nielsen-Olesen-like solitons of this action minus
the current term.   
The classical soliton is a sum (with given collective coordinates)
of vortex terms of the form:
\begin{equation}
\label{abgaugepot}
  A_i(x)= \pm\sum_V \pi \oint_{V}dz_k\epsilon_{ijk}\partial_j
[\Delta_m(x-z)-\Delta_0(x-z)] \equiv U_i. 
\end{equation}
$\Delta_{m,0}$ is the free mass-$m$ or massless propagator, the sum is over
a
condensate of very long (compared to the persistence length; see
below) closed loops labeled by $V$ that are mutually- and self-avoiding. 
The   confining part of the solitonic gauge potential comes solely from the
$\Delta_0$ term, which is a singular pure-gauge term coming from the
Goldstone-like field $\phi$. (The short-distance Dirac-string singularities
cancel between the two terms.)  For purposes of studying the area law only
the $\Delta_0$ term need be saved in Type I gluons.

In addition, there is a Type II term in the vector potential coming from
gluons having the Wilson-loop current as source: 
\begin{equation}
\label{massterm}
\Delta_mJ_i\equiv \int (\nabla^2+m^2)^{-1}J_i.
\end{equation} 
 The relevant terms in the action that involve this loop are:
\begin{equation}
\label{action}
I_J= \int \{-U_iJ_i+\frac{1}{2}J_i\frac{1}{\nabla^2+m^2}J_i   \};
\end{equation}
the first term in brackets is Type I, and the massive term is Type II.  The
Type II term is conventional and gives rise to a perimeter term.  In our
Abelian model the Type II gluons do not interact with each other, but it is
essential for the L\"uscher term that they do interact, as indeed they do
in a non-Abelian gauge theory.

As
the collective position coordinates
of each loop $V$ vary, an ensemble of vortices is realized.       For
$SU(2)$
quarks, the  VEV $\langle W\rangle$ of the Wilson loop $\Gamma$ is the
ensemble average:
\begin{equation}
\label{wvev} 
\langle W\rangle =\langle \exp [i  \oint_{\Gamma}dz_iA_i(z)]\rangle .
\end{equation}
Insert the $\Delta_0$ term of Eq.~(\ref{abgaugepot}) to find:
\begin{equation}
\label{linkform}
\langle W\rangle \equiv \langle \exp [i\pi\sum_VLk_V]\rangle
\end{equation}
where $Lk_V$ is the Gauss link number of vortex $V$ with the Wilson loop, as
given in the standard integral:
\begin{equation}
\label{lkstand}
Lk_V \equiv \oint_{\Gamma}dx_i \oint_{V}dz_k\epsilon_{ijk}\partial_j
\Delta_0(x-z).
\end{equation}
Consequently, $\langle W\rangle$ is simply the ensemble average of a product
of -1's, with a -1 for every piercing by a vortex of any surface spanning
the Wilson loop.  

This is an area law, as the well-known $d=2$ derivation shows.  To implement
a finite persistence length $\lambda$ put the vortices on a lattice of unit
length $\lambda$, and the flat Wilson loop on a coordinate plane of the
dual lattice. That portion of the coordinate plane bounded by the Wilson
loop is a flat surface spanning the Wilson loop; it is pierced at places by
vortices.    To implement self-avoidance we take it that at most one vortex
can pierce any unit square of this spanning surface, and we assign a
probability $p$ that such a unit square is in fact pierced by a vortex.
There is an areal density $\rho$ of vortex pierce points on the flat
surface, equalling $p/\lambda^2$.
Assume for simplicity that all vortex piercings are uncorrelated and that
every piercing of a Wilson loop is equivalent to a contribution $\pm$1 to
the VEV (which is not true, but taking this into account only changes the
value of, not the existence of, the string tension; see \cite{corn135}).  It
is
then easy to see that:
\begin{equation}
\label{flatloop}
\langle W\rangle =(\bar{p}-p)^N =\exp [\ln (1-2p)A/\lambda^2]
\equiv e^{-\sigma_0A}
\end{equation}
where $\bar{p}=1-p$ is the probability that a square of the spanning surface
is not pierced by a vortex (thus giving a factor of 1 in the VEV) and
$N=A/\lambda^2$ is the number of lattice squares in the spanning surface,
of area $A$.  

In this $d=2$ case it is pretty obvious what surface and what area is going
to show up in the Wilson-loop VEV, but what about $d>2$ and non-planar
Wilson loops?  It is true that
by Stokes' theorem the Gauss link number can be displayed as an integral
showing the piercing by a vortex of {\em some} surface spanning the Wilson
loop, but {\em any} surface will do, since the link number is purely
topological.  Yet the law of confinement is expressed solely in terms of the
Wilson loop contour and must make reference to a (generically) unique
surface for a given contour.  One can see semi-quantitatively
\cite{corn135} that the larger the area of a spanning surface the fewer
vortices are actually linked (the rest have, for $SU(2)$, an even Gauss
link number), which is to say that the probability $p$ of
Eq.~(\ref{flatloop}) depends on the surface and diminishes as the surface
area grows.   Maybe then it is plausible that the area to be associated
with linked vortices, in the sense of Eq.~(\ref{flatloop}), is a minimal
area.  But to my knowledge this has never been shown convincingly from
first principles.  

Since only Type I gluons contribute to the area law, it is much simpler to
simulate the area laws for non-planar Wilson loops because one need not
simulate the full dynamics of QCD.  All that is needed is to construct an
ensemble of random, self- and mutually-avoiding closed loops and to
calculate the area law and its fluctuations from topological formulas such
as Eq.~(\ref{linkform}).  Construction of such ensembles is well-known in
polymer physics [see, for example, \cite{kremer}].  It would be very
interesting to prove or disprove the conjecture that {\em purely
topological} confinement leads to a minimal surface and L\"uscher term. 
Equally fascinating is the study of area laws for more complex Wilson
loops, such as the change in the area law with $z$ of two coaxial loops
separated by a distance $z$ \cite{corn135}.

\section{Fishnet graphs and gluon chains }

There is no convincing first-principles explanation that I know telling us
why the area in the area law is like a physical membrane with a surface
tension and not just a mathematical area somehow defined by the topology of
a sea of vortices.  (Of course, there is a lot of lattice evidence and no
one doubts the reality of this surface tension.)  I will make some remarks
that probably are no more convincing than any other hand-waving
explanation, but which I feel are on the right track.  They have to do with
the old subject of fishnet graphs \cite{nielole,sakvira,thooft2}, and the  
less old subject of gluon chains \cite{tikt,greenthorn,greenole}.  One of
the motivating factors for gluon chains is that they lead to a picture
rather like that of string in string theory, and should naturally
accommodate a L\"uscher force and related phenomena.

\begin{figure}
\begin{center}
\includegraphics[width=3in]{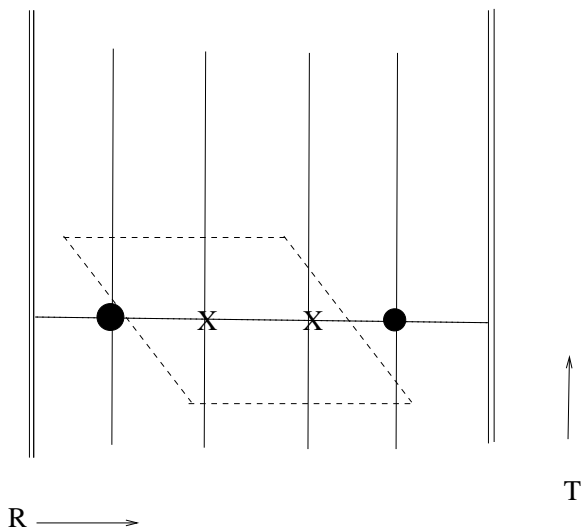}
\caption{\label{chain} Double lines are the edges of one Wilson loop. 
Dashed lines show a perpendicular Wilson loop.  Dotted lines with circles
and Xs show gluon chain in R direction, intersecting the dashed-line Wilson
loop at the Xs.  }
\end{center}
\end{figure}

Begin with gluon chains.  In a simplistic description of previous work, a
gluon chain joining a static quark and antiquark (which I think of as two
opposite sides of a large Wilson loop) consists of a string of localized
gluons joined to their nearest neighbors (including quarks) by a force that
increases no less rapidly than linear.  I show a sketch of the chain in
Figure 1.  In the figure the gluons are shown as localized to a particular
time, but of course they actually have world lines (the vertical lines in
the figure). The long string of gluon nearest-neighbor interactions, at
fixed time, reduces this growth to linear (or if the original gluon-gluon
force was linear, may reduce the string tension, as needed in Coulomb gauge
\cite{greenole}).  Since the gluons are supposed to be massless, some sort
of unspecified localization mechanism has to exist.  Moreover, it is not
clear what exactly gives the linear or super-linear force law between
nearest-neighbor gluons in the first place.  

There is a kind of {\em dual} to the gluon-chain model in which gluons are
replaced by nexuses\footnote{Nexuses are strictly non-Abelian and cannot
exist in isolation, but only as parts of center vortices; when nexuses are
included, center vortices are much more complex objects than the simple
Nielsen-Olesen vortex.} (essentially chromomagnetic monopoles) joined by
chromomagnetic flux tubes, whose energy naturally grows linearly with
separation \cite{ambgreen,corn122,corn123}.  If a gluon has dynamical mass
$m$, a nexus has mass of order $4\pi m/g^2$, where $g^2/(4\pi )\sim
\mathcal{O}(1)$ is the QCD running coupling in the infrared.  In fact, a
center vortex   in $d=3(4)$ is a closed string (two-surface) carrying the
chromomagnetic flux of the nexuses, which, as point-like (world-line)
objects,  divide the string (surface) into domains of differing flux
orientations.  There are also center vortices, with a single orientation,
with no nexuses.\footnote{The linkage of ordinary center vortices with
nexus world lines is one way of producing non-integral topological charge
\cite{corn113,engelrein,corn125}.}  In $d=4$ the closed center-vortex
two-surface can terminate on an 't Hooft loop.  'T Hooft's confinement
criterion \cite{thooft} says that if there is electric confinement (an area
law for the Wilson loop) there is no confinement for the 't Hooft loop. 
The way it comes about is that electric confinement requires a dominant
entropy contribution to the center-vortex action, in which case the center
vortices, or their constituents the nexus chains, tend to be very long and
space-filling.  In such a case the magnetic ``quarks'' in the 't Hooft loop
feel no long-range force, just as if two of them were joined by a floppy
string very long compared to the quarks' separation.  Conversely, there is
no entropy-driven $d=3,4$ condensate of ordinary gluons in QCD.  

Another statement of this duality is to think of the gluon chains as a $d=2$
``condensate", which needs a source---a
Wilson loop.  This condensate can also be described as a fishnet graph,
with massive gluons.    The mass provides a localization mechanism for the
gluons, which we can idealize as non-relativistic.  So the mass is also the
dominant source of string energy.  
In the (purely hypothetical) limit of chromoelectric-magnetic duality, the
properties of the $d=2$ gluon-chain condensate should be much like the
properties of a $d=2$ slice of the space-filling vortex condensate living
in $d=3,4$.  I have already reduced this vortex condensate to $d=2$ terms
in the discussion of topological confinement, with an areal density $\rho_M
=p/\lambda^2$ (the subscript $M$ indicating that the condensate is of
chromomagnetic vortices).  Equate this to the intergluon density $\rho_E$
of the gluon chain model, where $\rho_E\approx \zeta^2$, in terms of  the
intergluon spacing  $\zeta$.  The gluons have mass $m$, so the energy of a
gluon chain of length $R$ is
\begin{equation}
\label{chainenergy}
\epsilon = \sigma R \approx m\zeta R.
\end{equation}
Earlier I argued that $\sigma \approx 2\zeta^2$, so $\zeta\approx m/2$,
$\sigma \approx m^2/2$.  This happens to work fairly well for $m\approx$
600 MeV.

\begin{figure}
\begin{center}
\includegraphics[width=2.3in]{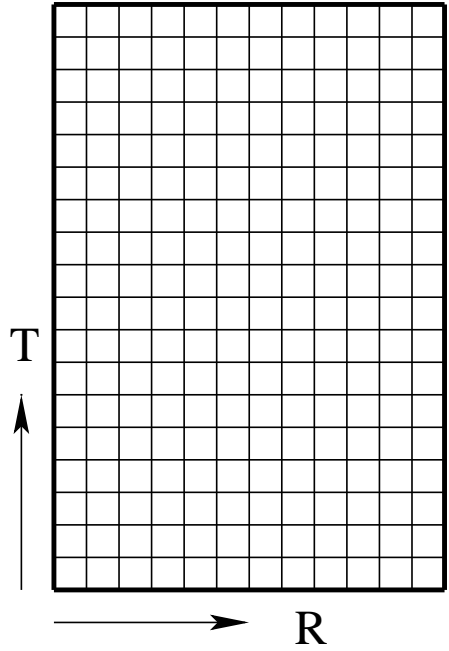}
\caption{\label{fishnet}  A fishnet graph in a Wilson loop (heavy lines). 
Vertices  labeled by coordinates
$Z_{\beta}(\sigma_1,\sigma_2)$; lines point along unit
vectors $\hat{e}(R,T)$   }
\end{center}
\end{figure}

Next, the question of the L\"uscher term, which I discuss in very sketchy
terms.  Figure \ref{fishnet} shows a Wilson loop decomposed as a fishnet
graph; for simplicity, I deal only with four-gluon interaction terms
explicitly  and treat gluons as massive scalar particles.  Of course, the
fishnet graph is not this regular lattice, since all vertex coordinates
must be integrated over. I assume that the average distance between
four-gluon vertices is of the order of $\lambda$, the intervortex distance
introduced earlier.  Let $\sigma_1$ be a coordinate in the plane along the
$R$ direction, and $\sigma_2$ a coordinate in the $T$ direction.  Every
four-gluon vertex has $d=4$ coordinates $Z_{\beta}(\sigma_1,\sigma_2)$.  
As in \cite{sakvira} write the value $G$ of the fishnet graph as:
\begin{eqnarray}
\label{sakvira}
G & = & const.\times\prod_{\sigma}\{d^4Z_{\beta}(\sigma)  
\Delta [(x-y)(\sigma_a)]\}\\ \nonumber
& = & const.\times \int (d^4Z_{\beta})\exp \{\sum_{\sigma} \ln \Delta
[(x-y)(\sigma_a)] \}.
\end{eqnarray}
Approximate the propagation distances $x-y$ by the first term in a
derivative expansion, so that with:
\begin{equation}
\label{logprop}
\ln \Delta [(x-y)^2]=\int_0^{\infty} \frac{ds}{s}\int_x^y(dz)
\exp \{-\frac{m}{2}\int_0^sd\tau [ \dot{z}^2 +1]\} 
\end{equation}
(dots indicate proper-time derivatives)
and the expansion:
\begin{eqnarray}
\label{fishcoord}
x_{\beta} & = & Z_{\beta}(\sigma_a+\lambda \hat{e}^{R,T}_a),
y_{\beta}=Z_{\beta}(\sigma_a)\\ \nonumber 
x_{\beta}-y_{\beta} & \approx & \lambda \hat{e}(R,T)\cdot\partial Z_{\beta}
(\sigma )\\ 
\end{eqnarray}
(where, for example, $\hat{e}(R)$ is a unit vector in the $R$ direction).
I further assume that the proper-time integral over $s$ can be replaced by
multiplication by $\Delta s\approx \lambda \approx 1/m$.
Sum over unit vectors to find the fishnet graph value roughly:
\begin{equation}
\label{roughgraph}
G\approx \int (dZ_{\beta}) \exp \{-\int \frac{d^2\sigma}{\lambda^2}
\frac{m}{2\Delta s}[(\lambda \partial_aZ_{\beta})^2]\}.      
\end{equation}

\begin{figure}
\begin{center}
\includegraphics[width=3in]{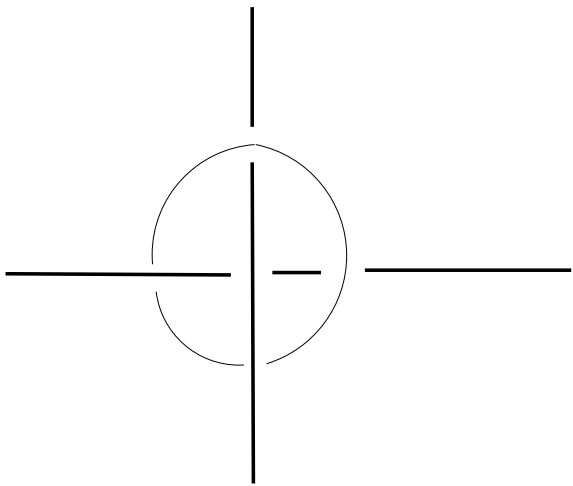}
\caption{\label{vertex}   Heuristics of vertex equilibria:  Like two strings
held together by a frictionless ring.  They can slide freely, but they can't
separate}
\end{center}
\end{figure}

The semiclassical approximation to $G$ yields the equation
$(\partial_a)^2Z_{\beta}=0$.  We interpret this as coming from a
force-balance relation at every vertex, with the gluon forces along the
world lines meeting at the vertices.  If two world lines are in the $R$
direction and the other two in the $T$ direction this force balance reads:
\begin{eqnarray}
\label{forcebal}
Z_{\beta}(\sigma_1+\lambda,\sigma_2 )-Z_{\beta}(\sigma_1,\sigma_2)
& + &Z_{\beta}(\sigma_1-\lambda,\sigma_2 )-Z_{\beta}(\sigma_1,\sigma_2) \\
\nonumber
+{\rm two\;other\;terms}=0 & \rightarrow & \lambda^2
(\partial_a)^2Z_{\beta}=0.  
\end{eqnarray}
However, this specification is incomplete, since the world lines need not
point along the coordinate axes and they can join anywhere.  The physical
situation is something like what is shown in Figure \ref{vertex}, showing
two frictionless strings bound at one point by a frictionless ring that
allows sliding but not separation of the strings.
A further specification of force balance is:
\begin{equation}
\label{slide}
\partial_a Z_{\beta}\partial_bZ_{\beta}=0\;\;(a\neq b)
\end{equation}
and isotropy demands:
\begin{equation}
\label{isotropy}
\partial_1 Z_{\beta}\partial_1Z_{\beta}=  \partial_2
Z_{\beta}\partial_2Z_{\beta}\;\;[=f^2(\sigma )].
\end{equation}
In other words the induced metric is in conformal gauge:
\begin{equation}
\label{confgauge}
\eta_{ab}=\partial_a Z_{\beta}\partial_bZ_{\beta}=\delta_{ab}f^2(\sigma ).
\end{equation}

Now there are many minima of the fishnet action of Eq.~(\ref{sakvira}) if no
further conditions are imposed, but if the conformal gauge condition is
imposed, the action is the Dirichlet action of a minimal surface. The
force-balance conditions used here support the obvious conjecture that a
fishnet graph indeed corresponds to a surface under tension, with as usual
a L\"uscher term.

\section{A return to gauge invariance on the lattice?}

There is by now a vast body of lattice simulations done in the last few
years
in particular gauges, notably the Landau gauge.  This seems to violate the
basic premise  
of lattice gauge theory, which is gauge invariance; no gauge-fixing terms
were needed to define on-shell functional integrals.  The apparent price to
pay
for gauge invariance was that we were forbidden to look at off-shell Green's
functions.   And if a gauge is fixed, allowing off-shell calculations, the
Green's functions were not gauge-invariant and hence unphysical. 
Nevertheless there is good reason for lattice theorists and simulators to
do gauge-fixed
simulations, because otherwise the toolset for understanding the mechanisms
of QCD is too limited.  Only with off-shell Green's functions can one
construct and solve Schwinger-Dyson equations (SDEs), to understand, for
example,  how gluons get a dynamical mass and many other issues.

   In the continuum we have the pinch
technique (PT) \cite{corna7,corna9,corn076} that showed how to extract
off-shell Green's functions that really were gauge-invariant, at least at
the one-loop level.  In a general gauge this requires a complicated
recombination of standard Feynman graphs and graph pieces.   Years later,
Papavassiliou and Binosi (who will tell
you more about it in this Workshop) made the ultimate advance, showing
(after some suggestive one-loop work by others) that
the all-order generalization of the PT was nothing but good old Feynman
graphs calculated in the background-field Feynman gauge \cite{binpap}; no
recombination necessary in this gauge.

   It is perhaps ironic that the way to find PT Green's functions is to
choose a particular gauge---the background-field Feynman gauge.  The point
is, of course, not that {\em a priori} specification of this gauge is
particularly important, but that the gauge-invariant off-shell Green's
functions of the PT happen to be those calculated in this gauge.

There are at least two ways to do the PT rearrangement on the lattice.
\begin{enumerate}
\item Translate into lattice language the steps used in the continuum PT, in
which parts of standard
Green's functions in {\em any} gauge, for example Landau gauge, are
recombined, using Ward identities and other special tools, and
added together in gauge-invariant combinations.    
\item   Make use of the work of Dashen and Gross \cite{dashg} and of
Cucchieri {\em et al.} [\cite{cucch} and this Workshop] to find out how to
implement the background-field Feynman gauge in lattice simulations.
\end{enumerate}
Neither of these approaches is easy, or they would be done by now. 

The first way is very complex, but may be useful because it can start from
the Landau gauge, which is by now well-understood on the lattice.   As for
the second way,  Dashen
and Gross \cite{dashg} long ago showed how to implement gauge-fixing to the
background-field Feynman gauge on the lattice, but only in lowest-order
perturbation theory.   Cucchieri and collaborators [see \cite{cucch} and
references therein] spent considerable effort trying to generalize the
usual Landau gauge-fixing algorithm, which is in fact a minimization over
gauge transformations, to general covariant gauges as specified by the
standard $\xi$-parameter ($\xi=0$, Landau gauge; $\xi=1$, Feynman gauge). 
But for $\xi\neq 0$ there is  no simple minimization procedure that
generalizes Landau gauge.   Cucchieri {\em et al.} have recently found
\cite{cucch} a new formulation of $R(\xi )$ gauges that, while not trivial
to implement on the lattice, is more promising for numerical implementation
than earlier efforts.  Since Cucchieri is talking at this Workshop on this
subject I will not discuss it further here, but just comment that to me, at
least, there seems no particular extra difficulty in principle to extend
the work in \cite{cucch} to the background-field Feynman gauge, using the
work of Dashen and Gross. 

I urge the community of lattice theorists and simulators to think about
this:  How can you construct gauge-invariant off-shell Green's functions
for the lattice, and calculate their properties?

\end{document}